\documentclass[prb,twocolumn,showpacs,fleqn]{revtex4}

\usepackage{graphicx}
\usepackage{amsmath}
\usepackage{amsfonts}
\usepackage{amssymb}
\usepackage{epstopdf}
\usepackage{grffile}

\newcommand{\mean}[1]{\left\langle #1 \right\rangle}

\renewcommand{\imath}{\mathrm{i}}

\newcommand{\eps}{\varepsilon}

\usepackage{color}\definecolor{gray}{rgb}{0.5,0.5,0.5}

\def\nQD{n}
\def\eQD{\varepsilon}
\def\phivec{\mathbf{p}}
\def\phicom{p}

\def\tau{\phi}
\def\pperp{r}

\def\Int#1#2#3{\int_{#1}^{#2}\!\!d#3\,}

\def\rev{*}

\begin{document}
\title{Full Counting Statistics of a Non-adiabatic Electron Pump}
\author{Alexander Croy}
\email{croy@pks.mpg.de}

\author{Ulf Saalmann}
\affiliation{Max-Planck-Institute for the Physics of Complex Systems,
   N\"{o}thnitzer Str.~38, D-01187 Dresden, Germany}

\date{\today}


\begin{abstract}\noindent
Non-adiabatic charge pumping through a single-level quantum dot with periodically modulated parameters is studied theoretically. 
By means of a quantum-master-equation approach the full counting statistics of the system is obtained. 
We find a trinomial-probability distribution of the charge transfer, which adequately describes the reversal of the pumping current by sweeping the driving frequency.
Further, we derive equations of motion for current and noise, and solve those numerically for two different driving schemes.
Both show interesting features which can be fully analyzed due to the simple and generic model studied. 
\end{abstract}

\maketitle

\section{Introduction}
Pumping of electrons through nano devices by a time-dependent modulation of
device parameters has received a lot of interest over the past years. Such pumps are
interesting for many applications, but they are particularly useful in metrology\cite{fl04,ke08}.
In this context, experimental and technological progress has lead to high-frequency
(in the GHz regime)\cite{blka+07,funi+08,kaka+08,giwr+10} and high-accuracy charge pumping\cite{gika+12}.

Moreover, pumping is also interesting for addressing fundamental questions connected with the transport
of quantum particles. Knowledge about the
full counting statistics (FCS) of the pumped electrons allows a detailed understanding
of relaxation and quantum effects\cite{emag11}. FCS for Coulomb-blockade systems has been studied theoretically in the context
of stationary\cite{bana03,no13}, driven\cite{ivle+97,abiv08,esha+09} and nano-electromechanical\cite{pi04,flno+05,becl08} 
systems. In particular, understanding the noise of the pumping current is relevant for high-accuracy pumping.
Accordingly, the noise in different setups of driven devices has been investigated theoretically \cite{lele94,cale+03,wuti10} and measured, for example, in a charge pump\cite{maho+08}.

While adiabatic pumping is very well studied\cite{br98,zhsp+99,enah+02,mobu01,cago+09,risp+13}, the description of non-adiabatic effects remains challenging.
In view of the experimental developments towards higher frequencies, this regime becomes increasingly
relevant\cite{kaka14}. Moreover, non-adiabatic driving can lead to interesting effects\cite{cago+09,crsa+12}, like
the possibility to reverse the pumping current by sweeping the driving frequency\cite{crsa12}.
In order to exploit such effects a better understanding of the FCS for fast pumping is necessary. 

In this article we consider an electron pump modeled by a single-level quantum dot at zero bias. Based on the formalism given before\cite{pi04} we calculate the cumulant generating function at zero temperature and for arbitrary driving schemes. We show that in the adiabatic regime, the probability distribution for pumping electrons is always
binomial, which implies that the pumping is unidirectional. In the non-adiabatic regime we obtain a trinomial
distribution, which also explains the occurrence of the current reversal\cite{crsa12}.
We show that fluctuations at the reversal may show either a minimum or a maximum, depending on the driving scheme.

The outline of the article is as follows. In the next section we derive the FCS for a quantum-dot
electron pump. We also discuss the calculation of the
pumping-current noise. In Sec.\ \ref{sec:results} we consider a specific time-dependence of
the tunneling rates and the dot energy and compare the numerical results for this model to the analytic
expressions obtained in Sec.\ \ref{sec:theory}. Finally, we conclude with a summary and a discussion.

\section{Theory}\label{sec:theory}
We consider a resonant-level model \cite{crsa+12} characterized by a time-dependent energy level $\varepsilon(t)$ and time-dependent couplings to the left and right reservoirs given by tunneling rates $\Gamma_{\rm L}(t)$ and $\Gamma_{\rm R}(t)$, respectively.
As indicated these parameters explicitly depend on time. The actual time dependence will be specified below. 

In order to obtain the FCS we consider the number of electrons $N$ tunneling through the left barrier.
Note, that in general this number is different for the right barrier. However, time-averaged quantities in the steady-state regime do not depend on the barrier (left or right) they are calculated for.
The statistics of $N$ can be found from the characteristic function
$\Phi(\chi,t)\equiv\mean{\exp[\imath \chi N]}$, where $\chi$ is the counting variable. 
Knowing $\Phi(\chi,t)$, one can deduce the moments $\mean{N^m(t)}$ by differentiation.

Under the usual conditions\cite{grde92}, short reservoir correlation-time and Coulomb blockade, the characteristic
function for the present model can be determined from the following equation\cite{bana03}
\begin{subequations}\label{eq:genfunc_eq}\begin{widetext}
\begin{equation}
	\frac{\partial}{\partial t}
	\left(
	\begin{array}{c}
		\phicom_0(t) \\ \phicom_1(t)
	\end{array}
	\right)
	=
		- \mathbb{L}_\chi (t)
		\left(
	\begin{array}{c}
		\phicom_0(t) \\ \phicom_1(t)
	\end{array}
	\right)
	\quad\mbox{ with }
	\mathbb{L}_\chi (t)\equiv
	\left(
	\begin{array}{cc}
		f(t) \Gamma(t) & -\bar{f}(t)\left[e^{-\imath \chi} \Gamma_{\rm L}(t) {+} \Gamma_{\rm R}(t)\right] \\
		-f(t)\left[e^{\imath \chi} \Gamma_{\rm L}(t) {+} \Gamma_{\rm R}(t)\right] & \bar{f}(t) \Gamma(t)
	\end{array}
	\right)\;,
\end{equation}
\end{widetext}
where we have used
\begin{align}
f(t) &\equiv \frac{1}{1+\exp(\eQD(t)/k_{\rm B}T)},\quad \bar{f}(t) \equiv 1-f(t), \\
\Gamma(t) &\equiv \Gamma_{\rm L}(t) + \Gamma_{\rm R}(t).
\end{align}
\end{subequations}
Note that the counting field $\chi$ ``measures'' tunneling from and to the left reservoir only.
For $\chi=0$, Eq.\,\eqref{eq:genfunc_eq} reduces to the usual (Markovian) master equation for the diagonal elements of the reduced density matrix\cite{gupr96}. In this case, the components of the vector $\phivec=(\phicom_0,\phicom_1)^{\rm t}$ correspond to $\phicom_0(t)=1-\nQD(t)$ and $\phicom_1(t)=\nQD(t)$ with $n(t)$ the average occupation of the level at time $t$.
The characteristic function is given in terms of the solution $\phivec$ of Eq.\,\eqref{eq:genfunc_eq} 
\begin{equation}
	\Phi(\chi,\tau) = \mathbf{q}\,\phivec(\chi,\tau) \quad\mbox{with }\mathbf{q}=(1,1).
\end{equation} 
The product is to be understood as a scalar product of two two-component vectors.

Provided, that all external parameters change periodically with frequency $\Omega$, we can substitute $\tau=\Omega t$ to get\cite{pi04}
\begin{equation}
	\frac{\partial}{\partial \tau}
	\phivec(\chi,\tau)
	= -\frac{1}{\Omega} \mathbb{L}_\chi (\tau)\phivec(\chi, \tau)
\end{equation}
with $\mathbb{L}_\chi$ defined in Eq.\ \eqref{eq:genfunc_eq}. The solution after one cycle can be written as 
$\phivec(\chi,\tau\,{+}\,2\pi)=\mathbb{A}\,\phivec(\chi,\tau)$ with the matrix
\begin{equation}\label{eq:Amatrix}
	\mathbb{A} \equiv \mathcal{T}\exp\left( -\frac{1}{\Omega}\int^{2\pi}_{0}\!\!\!d\tau\; \mathbb{L}_\chi (\tau)\right),
\end{equation}
where $\mathcal{T}$ denotes the time-order prescription.

Assuming that the counting fields can be switched on and off adiabatically, the generating
function for $k$ counting cycles is obtained from \cite{bana03,pi04}
\begin{equation}\label{eq:GFfromMatrix}
	\Phi_k = \mathbf{q}\, \mathbb{A}^k\,\widetilde{\phivec}\;,
\end{equation}
where $\widetilde{\phivec}$ is the steady-state solution for $\chi=0$. For a large number of cycles $k$
the characteristic function is determined by the largest eigenvalue \cite{bana03,pi04} of $\mathbb{A}$.

\subsection{Current and Noise}
For many applications it is sufficient to know the first two moments, $Q=\mean{N}$ and $\Delta Q^2=\mean{\Delta N^2}$,
of the pumped charge. The moments are given in units of $e$ and $e^2$. In the following, we will first derive a set of equations, which allow the calculation of both moments for arbitrary frequencies and temperatures. Since we assumed the counting to occur in the left reservoir, all time-dependent
quantities refer to that reservoir and the label is suppressed.

As will be shown, it is more convenient to consider the time-derivatives, $I(t)=\partial_t \mean{N}$ and
$S(t) = \partial_t \mean{\Delta N^2}$.
Starting from the definition of the current, one gets the following expression
\begin{widetext}
\begin{equation}
	I(t) = \frac{\partial}{\partial t} \mean{N}
					= \mathbf{q} \frac{\partial}{\partial t}\frac{\partial}{\partial (\imath \chi)}\phivec\Big|_{\chi=0}
					= -\mathbf{q} \frac{\partial \mathbb{L}_\chi}{\partial (\imath \chi)} \phivec\Big|_{\chi=0}
					= \Gamma_{\rm L}(t) \left[ f(t) - \nQD(t)\right]\,, \label{eq:current}
\end{equation}
where we have used Eq.\ \eqref{eq:genfunc_eq} and $\mathbf{q}\,\mathbb{L}_{\chi=0}=(0,0)$.
Similarly one finds
\begin{align}
	S(t) & = \frac{\partial}{\partial t} \left[\mean{N^2} - \mean{N}^2 \right]
					= \mathbf{q}\left[ \frac{\partial}{\partial t}\frac{\partial^2}{\partial (\imath \chi)^2}
					  - 2 I \frac{\partial}{\partial (\imath \chi)}\right]\phivec\Big|_{\chi=0}
					= -\mathbf{q} \left[ \frac{\partial^2 \mathbb{L}_\chi}{\partial (\imath \chi)^2}
					  - 2 \frac{\partial \mathbb{L}_\chi}{\partial (\imath \chi)}\frac{\partial}{\partial (\imath \chi)}
					  - 2 I \frac{\partial}{\partial (\imath \chi)}\right]\phivec\Big|_{\chi=0} \notag\\
				   &= - 2\Gamma_{\rm L}(t) \pperp (t) 
				      + \Gamma_{\rm L}(t) \left[\bar{f}(t)n(t)+f(t)\bar{n}(t)\right]\;,
				      \label{eq:noise}
\end{align}
\end{widetext}
where 
\begin{subequations}\begin{align}
\pperp(t) &\equiv \phicom'_{1}(t)  - \nQD(t)\left[\phicom'_{0}(t)+\phicom'_{1}(t)\right]\;,\\ 
\phicom'_{j}(t)&\equiv\frac{\partial \phicom_j(\chi,t)}{\partial (\imath\chi)} \Big|_{\chi=0}
\quad\mbox{for }j=0,1.
\end{align}	\end{subequations}
To calculate $I$ and $S$, the time evolution of the occupation $\nQD$ and the newly defined quantity $\pperp$ are needed. Both of them can be obtained from Eq.\ \eqref{eq:genfunc_eq}. The corresponding equations are
\begin{subequations}\label{eq:summary}
\begin{align}
	\frac{\partial}{\partial t} \nQD(t) = {} & 
	\Gamma(t) \left[ f(t) {-} \nQD(t) \right]	\;, \\
\frac{\partial}{\partial t} \pperp (t)
= {} & -\Gamma(t) \pperp (t)\\ \notag &
   +\Gamma_{\rm L}(t)\left[ f(t)\bar{f}(t)-[f(t){-}n(t)]^{2} \right]\,.\notag
\end{align}	
\end{subequations}
Equations \eqref{eq:summary} can be solved numerically in a straightforward way. The moments
${Q}$ and ${\Delta Q^2}$ are found by integrating $I$ and $S$ over one period, respectively. For
low frequencies and finite temperatures, one can obtain analytical expressions for both quantities.
This case is discussed in Sec.\ \ref{sec:lowfreqnoise} below.

\subsection{FCS at zero temperature}
In the following section we calculate the generating function at zero temperature,
i.e., when the Fermi function only attains the values $0$ and $1$.
To this end we assume that for $0\leq \tau < \pi$ the level is charged ($f{=}1$, $\bar{f}{=}0$) and
for $\pi \leq \tau < 2\pi$ it is de-charged ($f{=}0$, $\bar{f}{=}1$). In this case, the solutions to Eq.\ \eqref{eq:genfunc_eq} at the
end of the respective half-periods can be obtained analytically. Overall, we get for one period
\begin{widetext}
\begin{subequations}\label{eq:acdbcd-matrix}\begin{equation}
	\phivec(\chi,2\pi) = \mathbb{A}_{0}(\chi)\phivec(\chi,0)
	\quad\mbox{ with }
	\mathbb{A}_{0}(\chi)\equiv
		\left(
	\begin{array}{cc}
		1{-}\beta_{\rm L} {-} \beta_{\rm R} + (e^{-\imath \chi} \gamma_{\rm L} {+} \gamma_{\rm R})(e^{\imath \chi} \beta_{\rm L} {+} \beta_{\rm R}) 
		& e^{-\imath \chi} \gamma_{\rm L} {+} \gamma_{\rm R} \\
		(1 {-} \gamma_{\rm L} {-} \gamma_{\rm R}) (e^{\imath \chi} \beta_{\rm L} {+} \beta_{\rm R}) 
		& 1{-} \gamma_{\rm L} {-} \gamma_{\rm R}
	\end{array}
	\right)
\end{equation}
and the abbreviations
\begin{equation}
	\beta_\alpha \equiv \Int{0}{\pi}{\tau} \frac{\Gamma_\alpha(\tau)}{\Omega} \exp\left(- \frac{1}{\Omega}\Int{0}{\tau}{\tau'} \Gamma(\tau') \right)
	\,,\quad
	\gamma_\alpha \equiv \Int{\pi}{2\pi}{\tau} \frac{\Gamma_\alpha(\tau)}{\Omega} \exp\left(- \frac{1}{\Omega}\Int{\pi}{\tau}{\tau'} \Gamma(\tau') \right)\,.
\end{equation}\end{subequations}
\end{widetext}
Note that $\beta_\alpha$ and $\gamma_\alpha$ are completely determined by the time-dependence of the rates $\Gamma_\alpha$.
They define probabilities for charging or de-charging the level either from the left or the right reservoir, respectively.
Obviously $\beta_{\alpha},\,\gamma_{\alpha}>0$.
Further we note that
\begin{align}
	\beta_{\rm L}+\beta_{\rm R} 
	&=1-\exp\left(- \frac{1}{\Omega}\Int{0}{\pi}{\tau'} \Gamma(\tau') \right)<1.
	\label{eq:sum-beta}
\end{align}
Since both quantities, $\beta_{\rm L}$ and $\beta_{\rm R}$, are positive it follows from \eqref{eq:sum-beta} that $\beta_{\rm L},\,\beta_{\rm R}<1$.
Similar arguments hold for $\gamma_{\rm L}$ and $\gamma_{\rm R}$.
In order to simplify the notation, in the following we use the definitions
\begin{equation}
\bar{\beta}\equiv 1-\beta_{\rm L}-\beta_{\rm R},
\quad
\bar{\gamma}\equiv 1-\gamma_{\rm L}-\gamma_{\rm R}\;.
\end{equation}
Finally, the eigenvalues of the zero-temperature matrix $\mathbb{A}_{0}(\chi)$, which is defined in Eq.\,(\ref{eq:acdbcd-matrix}a), 
are given by
\begin{subequations}\label{eq:ExactGF}
\begin{align}
	\lambda_{\pm} &= \frac{1}{2}\left[g(\chi)\pm \sqrt{g^{2}(\chi) - 4 \bar{\beta}\bar{\gamma}}\right]\;,
	\\g(\chi) &\equiv \bar{\beta}{+}\bar{\gamma} + (e^{-\imath \chi} \gamma_{\rm L} {+} \gamma_{\rm R})(e^{\imath \chi} \beta_{\rm L} {+}	\beta_{\rm R})\;,
\end{align}\end{subequations}
from which the current and noise can be obtained by differentiation. 
The eigenvalue with the largest absolute value is $\lambda_{+}$. This expression is exact under the assumptions stated
above and can be used to obtain the FCS of the pumped charge.

To gain further insight about the nature of the statistics, it is useful to consider limiting cases.
If $\bar{\beta},\,\bar{\gamma} \ll \beta,\,\gamma$ one can write down the following generating function
\begin{align}
	(\Phi_k)^{1/k} \approx {} &\gamma_{\rm L}\beta_{\rm R} e^{-\imath \chi}
	+ \beta_{\rm L}\gamma_{\rm R} e^{\imath \chi}\notag\\
					&+ (\bar{\beta} + \bar{\gamma} + \beta_{\rm L}\gamma_{\rm L} + \beta_{\rm R}\gamma_{\rm R})\;.		
\label{eq:genfunlimit}\end{align}
This limit describes the situation where the level is almost fully charged in the first half cycle and correspondingly de-charged 
in the second half cycle. 
Note that the expression in Eq.\ \eqref{eq:genfunlimit} yields $\Phi_k\approx 1$ for $\chi=0$ only to first order in $\bar{\beta}$, 
$\bar{\gamma}$.

Equation \eqref{eq:genfunlimit} characterizes a trinomial probability distribution. Accordingly, there are three relevant probabilities 
\begin{equation}\label{eq:probe}
	p_- = \gamma_{\rm L}\beta_{\rm R},\ p_+ = \beta_{\rm L}\gamma_{\rm R}\ \text{and}\ p_0=1-p_--p_+,
\end{equation}
which describe a process of an electron being transferred to the left reservoir, to
the right reservoir or no transfer, respectively. It follows that the average charge and noise per period are
\begin{subequations}\label{eq:trinomial}
\begin{align}
	{Q} &= p_+ - p_-\;,\\
	{\Delta Q^2} &= p_+(1{-}p_+) + p_-(1{-}p_-) + 2 p_+ p_- \;. \label{eq:trinomial_dQ2}
\end{align}
\end{subequations}
For $p_+=p_-$ one finds that the pumped charge vanishes ${Q}=0$, while the noise
remains finite ${\Delta Q^2}=2p_+$ for $p_{+}{>}0$. At the vicinity of this point one obtains a
current reversal\cite{crsa+12,crsa12}.

In the opposite limit, $\bar{\beta},\,\bar{\gamma} \approx 1$, where the energy level is nearly empty during one cycle,
one can approximate the generating function by
\begin{equation}\label{eq:HighFreqGF}
	(\Phi_k)^{1/k} = c + \sqrt{(e^{-\imath \chi} \gamma_{\rm L} {+} \gamma_{\rm R})(e^{\imath \chi} \beta_{\rm L} {+}	\beta_{\rm R})}\;.
\end{equation}
Here, $c$ is independent of $\chi$ and guarantees $\Phi_k|_{\chi=0}=1$.
Equation \eqref{eq:HighFreqGF} describes in general a complicated probability distribution. However, if one barrier is
dominating the charging and the other one dominates the de-charging, one obtains a binomial distribution for a half-charge transfer.
For example, if $\beta_{\rm R}\approx\gamma_{\rm L}\approx 0$ one gets
\begin{equation}
	(\Phi_k)^{1/k} \approx c' + e^{\imath \chi/2} \beta_{\rm L}\gamma_{\rm R}\;.
\end{equation}
This ``fractional'' behavior has been discussed previously in the context of a nano-electromechanical charge shuttle\cite{pi04}.
It reflects the fact that the cycles are no longer independent of each other. For many counting cycles, however, the behavior
can effectively be described by the independent transfer of fractional charges. For later reference we note that the
charge per cycle can be obtained from Eq.\ \eqref{eq:HighFreqGF} to read
\begin{equation}
	{Q} = \frac{\beta_{\rm L}\gamma_{\rm R}-\beta_{\rm R}\gamma_{\rm L}}{2\sqrt{(\beta_{\rm L} {+}	\beta_{\rm R})(\gamma_{\rm L} {+}	\gamma_{\rm R})}}\;.
\end{equation}

\subsection{Noise for low frequencies and finite temperatures}\label{sec:lowfreqnoise}
At low frequencies, it is sufficient to consider the instantaneous contribution from
Eqs.\ \eqref{eq:summary}, as done similarly before\cite{risp+13}. Consequently, the time-derivatives are set to zero and one obtains
\begin{equation}
	\nQD(t) = f(t)\;,\quad
	\Gamma(t) \pperp(t) =
	\Gamma_{\rm L}(t) f(t) \bar{f}(t)\;.
\end{equation}
The instantaneous current and noise become
\begin{equation}
	I(t) = 0\;,\quad
	S(t) =  \frac{2 f(t) \bar{f}(t)}{1/\Gamma_{\rm L}(t)+1/\Gamma_{\rm R}(t)}  \;.
\end{equation}
Not surprisingly, the current is zero, since the level is at all times in equilibrium with the reservoirs. The noise is nonzero and is called equilibrium, or Nyquist-Johnson noise \cite{blbu00}. It is caused by the statistical nature of
the electron occupation in the reservoirs characterized by the Fermi distribution. Using the property
$f\bar{f} = -k_{\rm B} T \frac{\partial f}{\partial \varepsilon}$ the expression for the noise
becomes
\begin{equation}
	S(t) =  -2k_{\rm B} T \frac{1}{1/\Gamma_{\rm L}(t)+1/\Gamma_{\rm R}(t)} \frac{\partial f}{\partial \varepsilon}
		         \;.
\end{equation}
The noise per period is obtained from integrating over one period,
\begin{equation}
	\Delta Q^2 =
-2\frac{k_{\rm B} T}{\Omega} \int\limits^{2\pi}_0 d\tau\; 
\frac{[\partial \eQD(\tau)/\partial \tau]^{-1}[\partial f(\tau)/\partial \tau]}%
{1/\Gamma_{\rm L}(\tau)+1/\Gamma_{\rm R}(\tau)}
\label{eq:thermal_noise}
\end{equation}
The integral gives a value, which is independent of $\Omega$ but depends on the details of the external driving, i.\,e., the time-dependent level and tunneling rates.
Consequently, the noise per period will be proportional to
$k_{\rm B} T/\Omega$.

\section{Results}\label{sec:results}
\begin{figure}[b!]
\centering
	\includegraphics[width=0.9\columnwidth]{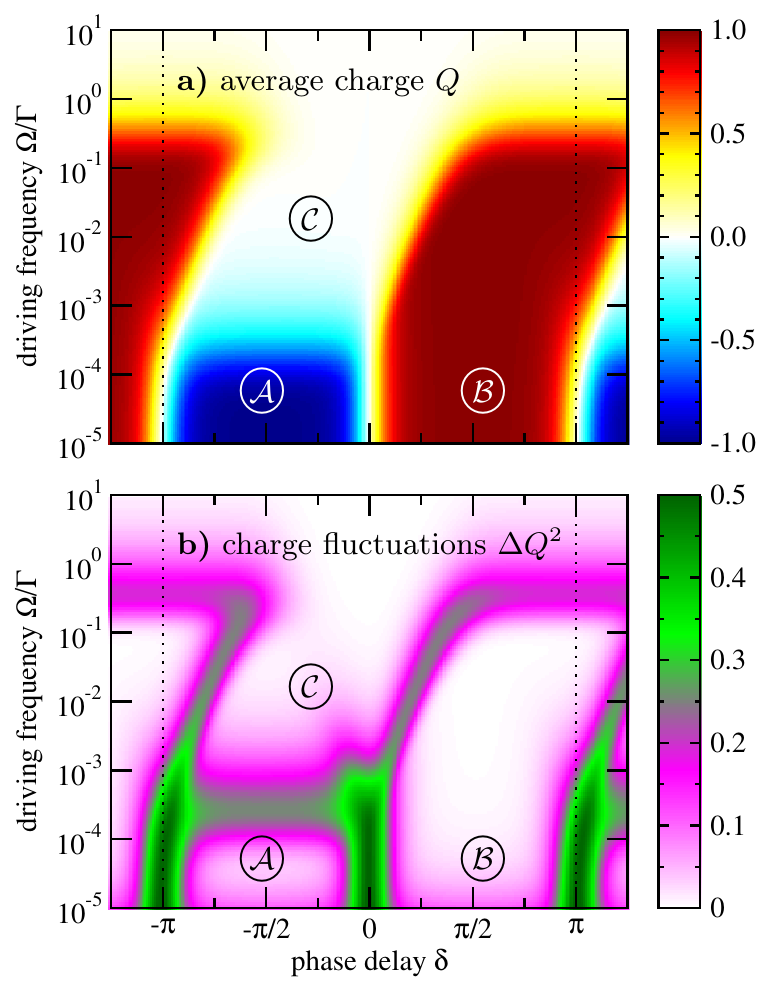}
	\caption{Pumped charge $Q$ and fluctuations $\Delta Q^{2}$ as a function of phase delay $\delta$ and driving frequency $\Omega$ 
	as obtained with the set of Eqs.\,\eqref{eq:current}\,--\,\eqref{eq:summary}
	for the time dependence of quantum-dot level $\eps$ and couplings $\Gamma_{\rm R,L}$ given in Eqs.\,\eqref{eq:system}.}
	\label{fig:charg-fluc-contour}
\end{figure}%

We study the first two cycle-averaged moments of the FCS: the charge $Q$ and its fluctuations $\Delta Q^{2}$ for an electron pump which shows current reversal and rectification effect in the non-adiabatic regime\cite{crsa12}.
They are obtained by integrating $I(t)$ and $S(t)$, as given by Eqs.\,\eqref{eq:current} and \eqref{eq:noise}, respectively, over one cycle under steady-state conditions, i.\,e., $\widetilde\phivec(\tau)=\mathbb{A}\,\widetilde\phivec(\tau)$ with $\mathbb{A}$ defined in Eq.\ \eqref{eq:Amatrix}.

\begin{figure*}[t!]
\centering
	\includegraphics[scale=0.55]{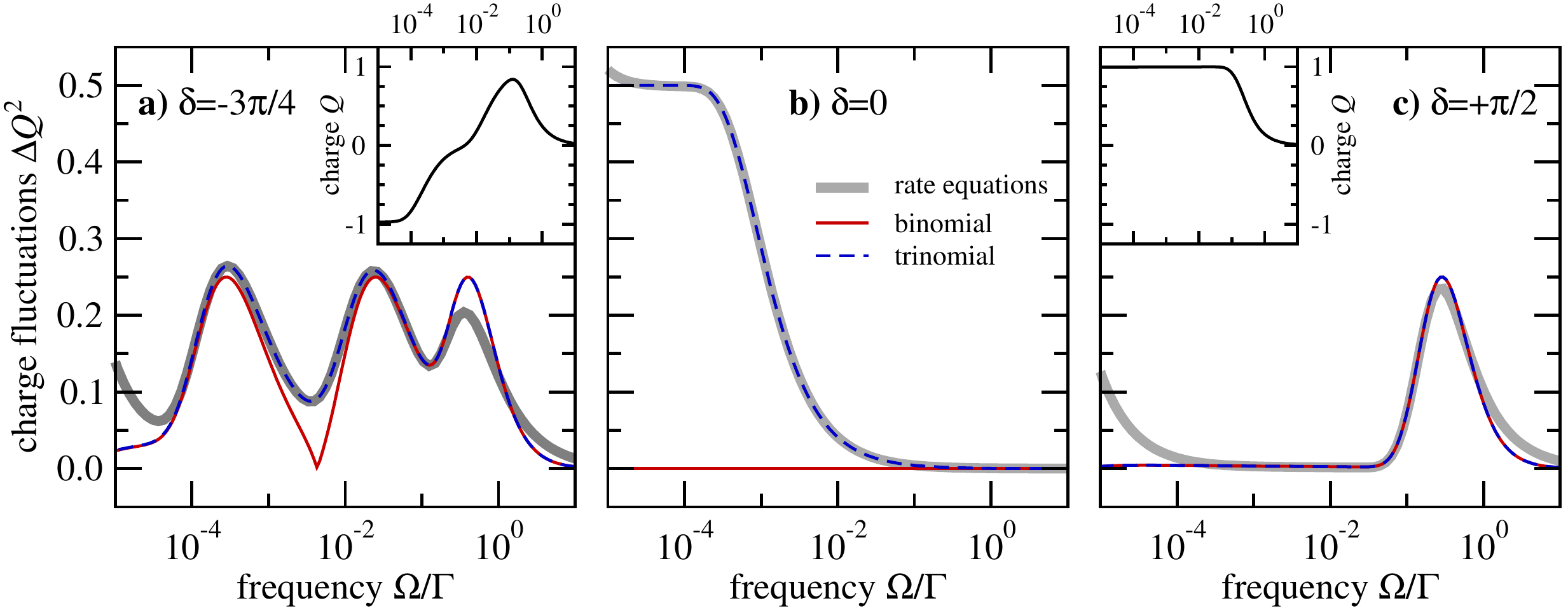}
	\caption{Charge fluctuations \eqref{eq:noise} obtained from the rate equations \eqref{eq:summary} for three selected phase delays $\delta$ and $k_{\rm B}T{=}\Gamma/100$. They are compared to those obtained for a trinomial probability distribution \eqref{eq:trinomial} with $p_- = \gamma_{\rm L}\beta_{\rm R}$, $p_+ = \beta_{\rm L}\gamma_{\rm R}$
	 and a binomial distribution function \eqref{eq:binomial} with $p\equiv|p_{+}{-}p_{-}|$.
	}
	\label{fig:charg-fluc-delays}
	\bigskip
	\includegraphics[scale=0.55]{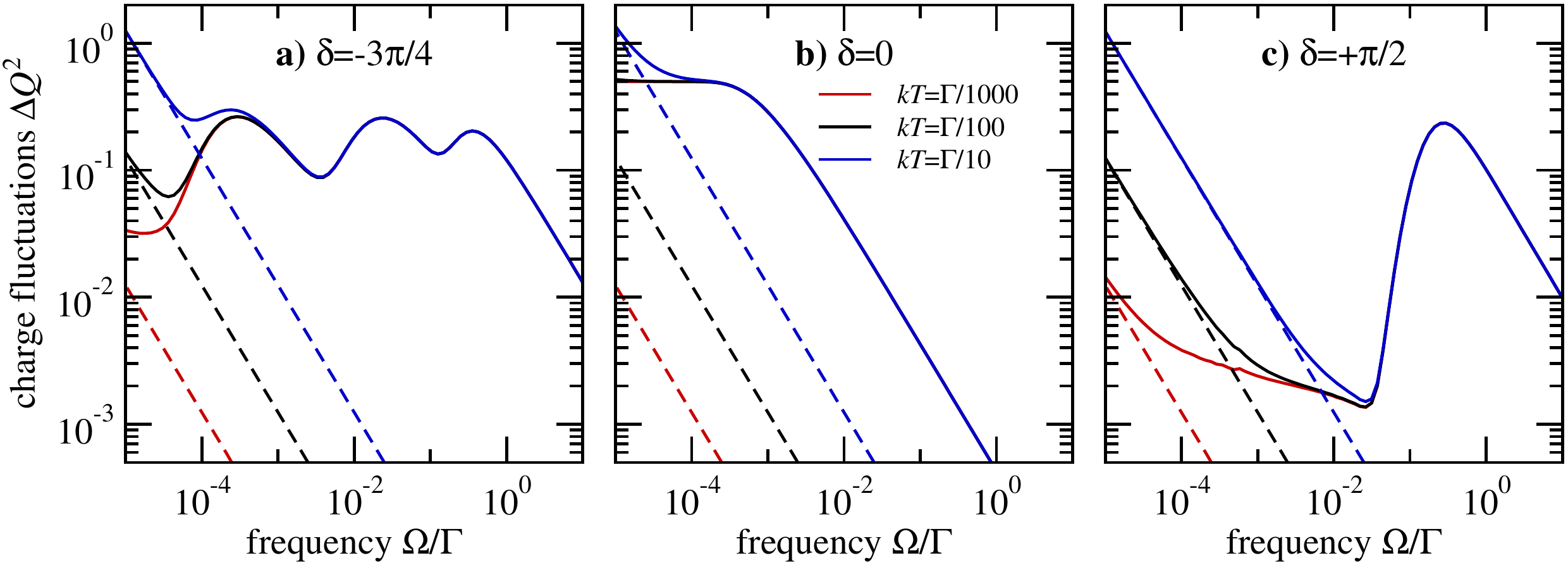}
	\caption{Charge fluctuations as shown in Fig.\,\ref{fig:charg-fluc-delays}, here with a double-logarithmic plot and for two additional temperatures $k_{\rm B}T=\Gamma/1000$ and $k_{\rm B}T=\Gamma/10$, respectively. The dashed lines show thermal noise according to Eq.\,\eqref{eq:thermal_noise}. 
	}
	\label{fig:charg-fluc-delays-logarithmic}
\end{figure*}%
\subsection{Exponential modulation}
To this end, Eq.\,\eqref{eq:summary} has been integrated numerically with the following pumping parameters
\begin{subequations}\label{eq:system}\begin{align}
\eps(t) & = \cos(\Omega t),\\
\Gamma_{\rm L}(t) & = [\Gamma/2]\exp\left(6[\cos\left(\Omega t-\delta\right)-1]\right),\\
\Gamma_{\rm R}(t) & = [\Gamma/2]\exp\left(6[\cos\left(\Omega t\right)-1]\right).
\end{align}\end{subequations}
The same parameters have been used before\cite{crsa12}.
They allow for the exponential dependence of tunneling rates due to oscillatory gate-voltages as used in experiments\cite{kaka+08,giwr+10}.
The pumping is characterized by two parameters, the driving frequency $\Omega$ and the time or phase delay $\delta$ of the left barrier with respect to the right one. The latter one is locked to the level $\eps$ which oscillates around the Fermi energy of both reservoirs $\mu_{\rm L}=\mu_{\rm R}=0$. 
If not specified differently, the temperature in the contacts is $k_{\rm B}T=\Gamma/100$.

Figure~\ref{fig:charg-fluc-contour} gives an overview of the results. 
We have varied the frequency $\Omega$ in a rather large range (10$^{-5}$\ldots\,10\,$\Gamma$) in order to cover the asymptotic adiabatic and non-adiabatic behavior for low and high frequencies, respectively.
One can clearly see (Fig.\,\ref{fig:charg-fluc-contour}a) that the direction of the charge transport does not only depend on the phase delay $\delta$ which controls the oscillation of the left barrier with respect to the right one.
Rather the frequency $\Omega$ is equally important and may be used to cause a current reversal as was discussed in detail before \cite{crsa12}.

The charge fluctuations (Fig.\,\ref{fig:charg-fluc-contour}b) show a similar pattern as the average charge, albeit somehow inverted.
Not surprisingly they vanish (white areas in Fig.\,\ref{fig:charg-fluc-contour}b), when the charge transport is \emph{quantized\/} either to the left (blue areas marked with $\cal A$ in Fig.\,\ref{fig:charg-fluc-contour}a) or to the right (red areas marked with $\cal B$ in Fig.\,\ref{fig:charg-fluc-contour}a) direction.
In regions, however, where this is not the case the fluctuations are finite, reaching sometimes a value of $\Delta Q^{2}=1/2$.

One maybe tempted to explain the charge fluctuations with the well-known expression \cite{bu90} 
\begin{equation}\label{eq:binomial}
\Delta Q^{2}=p\,(1-p),
\end{equation}
with $p$ the probability to transfer one charge per cycle. 
It is $p=|Q|=|p_{+}{-}p_{-}|$ with $p_{\pm}$ defined in Eq.\,\eqref{eq:probe}.
As we will see this binomial description is only applicable for finite values of $Q$.
It fails in regions where the current changes direction, i.\,e., where $Q{=}0$.
There are two exceptions where $Q$ and $\Delta Q^{2}$ vanishes simultaneously. 
One is the region marked with $\cal C$ in Fig.\,\ref{fig:charg-fluc-contour} and the other one is the high-frequency regime where $\Omega\gg\Gamma$. 
We will discuss all regions in the following.
To be more quantitative we have plotted the charge fluctuations for selected phase delays of $\delta{=}{-}3\pi/4$, $\delta{=}0$ and $\delta{=}{+}\pi/2$, respectively, in Fig.\,\ref{fig:charg-fluc-delays}.

\begin{figure*}[t!]
\centering
	\includegraphics[scale=0.55]{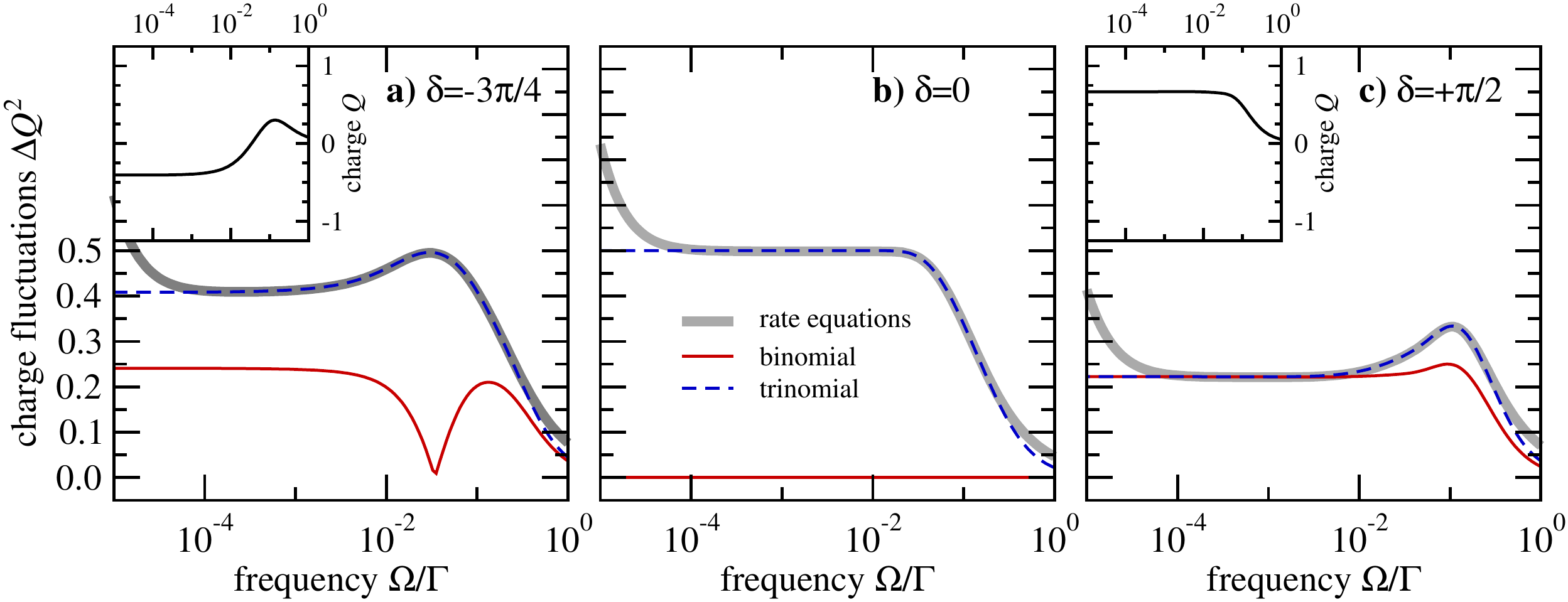}
	\caption{Charge fluctuations \eqref{eq:noise} for harmonic modulation \eqref{eq:hsystem}
        of the tunneling rates obtained from the rate equations \eqref{eq:summary} for three selected phase delays $\delta$ and $k_{\rm B}T{=}\Gamma/1000$. They are compared to those obtained for a trinomial probability distribution \eqref{eq:trinomial} with $p_- = \gamma_{\rm L}\beta_{\rm R}$, $p_+ = \beta_{\rm L}\gamma_{\rm R}$
	 and a binomial distribution function \eqref{eq:binomial} with $p\equiv|p_{+}{-}p_{-}|$.
	}
	\label{fig:harm-charg-fluc-delays}
\end{figure*}%
For $\delta{=}{-}3\pi/4$ there is a current reversal at $\Omega\approx5{\times}10^{-3}\Gamma$ (cf.\ inset in Fig.\,\ref{fig:charg-fluc-delays}a).
Along with the result from the rate equations (Fig.\,\ref{fig:charg-fluc-contour} and thick gray line in Fig.\,\ref{fig:charg-fluc-delays}) we show the binomial expression \eqref{eq:binomial} and the trinomial expression from Eq.\,(\ref{eq:trinomial}b).
The values of $p_{+}$ and $p_{-}$ in the latter case have been calculated numerically.
Note that this is an approximation since we have assumed $k_{\rm B}T{=}0$ in defining $p_{\pm}$.
Both shot-noise models agree qualitatively with the numerical result. 
However, at the current reversal ($Q{=}0$) charge fluctuations $\Delta Q^{2}$ are on the order of $1/10$, i.\,e., they do not vanish as predicted by Eq.\,\eqref{eq:binomial}.
The trinomial description (\ref{eq:trinomial}b) captures this behavior even quantitatively as Fig.\,\ref{fig:charg-fluc-delays}a shows.
Thus, the fluctuations reveal that the vanishing net transfer is not due to the fact that there is no transfer at all.
Rather transfer to the right and to the left do cancel each other exactly.

The discrepancy between binomial and trinomial description becomes even more obvious for $\delta{=}0$ which is shown in Fig.\,\ref{fig:charg-fluc-delays}b.
Due to symmetry ($p_{+}=p_{-}$) the charge $Q$ vanishes for all frequencies $\Omega$ and so does the binomial expression \eqref{eq:binomial}.
In contrast, the trinomial expression, which simplifies to $\Delta Q^{2}=2p_{+}=2p_{-}$, describes the fluctuations for all frequencies even quantitatively. 
Compared to the frequency-driven current reversal in Fig.\,\ref{fig:charg-fluc-delays}a the fluctuations are even larger here.

The case $\delta{=}{+}\pi/2$, shown in Fig.\,\ref{fig:charg-fluc-delays}c, is characterized by a quantized and thus fluctuation-free transfer over a wide range of frequencies ($\Omega\approx10^{-5}\ldots10^{-1}\Gamma$).

It remains to discuss the case where $Q$ and $\Delta Q^{2}$ vanish simultaneously. For large frequencies this is rather obvious, $p_{+}$ and $p_{-}$ become small for $\Omega\gg\Gamma$ since the transfer to either side can be neglected.
It is less clear for the region marked with $\cal C$.
It can be traced back to the coupling to right contact: both $\beta_{\rm R}$ and $\gamma_{\rm R}$, as defined in Eq.\,(\ref{eq:acdbcd-matrix}b), do vanish. 
If the level can neither be charged from the right ($\beta_{\rm R}$) nor de-charged to the right ($\gamma_{\rm R}$) the probabilities of transfer from ($p_{+}$) and to ($p_{-}$) the right contact do vanish as well.

Careful inspection shows that for all three cases the fluctuations increase above the values predicted by the two shot-noise models. 
The reason is that thermal or Nyquist-Johnson noise $\Delta Q^{2}_{\rm therm}$ starts to become larger than the shot noise.
This is shown in Fig.\,\ref{fig:charg-fluc-delays-logarithmic} where thermal contribution according to Eq.\,\eqref{eq:thermal_noise} is shown separately with dashed lines.
Clearly this occurs for larger temperatures $k_{\rm B}T$ at higher frequencies $\Omega$.
The frequency, at which this takeover occurs, depends on the phase delay $\delta$, as the three panels of Fig.\,\ref{fig:charg-fluc-delays-logarithmic} show.

\subsection{Harmonic modulation}
Taking, instead of (\ref{eq:system}b) and (\ref{eq:system}c),
\begin{equation}\label{eq:hsystem}
\Gamma_\alpha(t) = \Gamma_0 + \Gamma_1 \cos(\Omega t + \pi/2 - \delta_\alpha)
\end{equation}
yields a harmonic variation of the coupling rates, which is frequently used for modeling electron pumps.
The same setup has been investigated in view of a current reversal before\cite{crsa12}.
Using the same approach as in the last section, we calculate $Q$ and $\Delta Q^{2}$ for
$\Gamma_0=\Gamma_1=\Gamma/20$, $\delta_R=0$ and $\delta_L=\delta$. The results for a temperature of $k_{\rm B}T{=}\Gamma/1000$ are shown in
Fig.\ \ref{fig:harm-charg-fluc-delays}. 

In all three cases, the fluctuations obtained from the trinomial probability distribution \eqref{eq:trinomial} agree very well with the results of the rate equations \eqref{eq:summary}.
Comparing to Fig.\ \ref{fig:charg-fluc-delays} one sees that the results are qualitatively similar for $\delta=0$ and $\delta=\pi/2$. In the latter case and for harmonic modulation the fluctuations do not become
zero in the adiabatic regime since the pumped charge is not quantized (cf.\ inset of Fig.\ \ref{fig:harm-charg-fluc-delays}c) in contrast to the exponential driving (cf.\ inset of Fig.\ \ref{fig:charg-fluc-delays}c). 

The biggest qualitative difference is observed for $\delta=-3\pi/4$. For harmonic
modulation the charge fluctuations display a single maximum, which is attained in the vicinity
of the current reversal. 
This is in contrast to the exponential modulation, where the fluctuations
were found to have a minimum.
This difference is connected with the behavior of the left/right probabilities $p_{\pm}$ in the vicinity of the reversal frequency $\Omega_\rev$.
It turns out that the slopes have the same magnitude but different signs. The curvatures may be different.
Therefore we consider the following simple $\Omega$-dependence
\begin{equation}
    p_{\pm}(\Omega)=p_\rev\pm p'_\rev[\Omega{-}\Omega_\rev]+p''_{\rev\pm}[\Omega{-}\Omega_\rev]^{2}
\end{equation}
with $p_\rev=p_{\pm}(\Omega_\rev)$ and $Q(\Omega_\rev)=0$.
For such a para\-metrization the noise \eqref{eq:trinomial_dQ2} becomes up to 2nd order
\begin{equation}\label{eq:delta_min_max}
    \Delta Q^{2}(\Omega)\approx2p_\rev-[4p'_\rev{\!}^{2}{-}p''_{\rev}][\Omega{-}\Omega_\rev]^{2},
\end{equation}
where the abbreviation $p''_\rev\equiv p''_{\rev+}{+}p''_{\rev-}$ was used.
The noise has an extremum at $\Omega_\rev$, which can be a minimum or maximum depending on 
the relation of $4p'_\rev{\!}^{2}$ and $p''_{\rev}$.
Whereas in the harmonic case the curvatures of $p_{\pm}$ at $\Omega_\rev$ have different signs but similar magnitudes, and thus $p''_{\rev}\approx0$, the situation in the exponential case is different. Here one finds a non-linear increase/decrease of $p_{\pm}$ with $p''_{\rev}\gg4p'_\rev{\!}^{2}$. 
Correspondingly, close to the reversal we observe a maximum and a minimum of $\Delta Q^{2}(\Omega)$, respectively.

\section{Summary and Conclusions}
In summary, we have investigated the counting statistics for non-adiabatic
pumping of electrons through a single-level quantum dot. For zero temperature
we derived an analytic expression for the generating function [Eq.\ \eqref{eq:ExactGF}]
in terms of the probabilities for charging or de-charging the level during
one pump cycle. In the case where those probabilities are large (the level is 
almost completely filled and emptied), we found a trinomial probability 
distribution for the charge transfer. The associated elementary processes
correspond to an electron being transferred to the left reservoir, 
to the right reservoir or no transfer. This has the important consequence that
the transferred charge per cycle can vanish while the charge fluctuations remain
finite. It also shows that the current-reversal does not rely on interference 
effects.

Our findings are corroborated by numerical simulations of the first two
moments $Q$ and $\Delta Q^2$ for two driving schemes (exponential and harmonic). 
To this end we derived a set of ordinary differential
equations, valid for arbitrary time-dependencies, which were solved numerically. 
Those equations may also be used in connection with pulse-shaping techniques,
which allow for optimizing the pumping accuracy\cite{gika+12}.

Our calculations show that the driving frequency and the phase delay are important 
parameters, which both influence the \emph{statistics} of the pumping charge. This 
demonstrates that the ability to control the phase delays potentially provides an additional
knob to improve the performance of electron pumps.


\end{document}